# AI for Equitable Tennis Training: Leveraging AI for Equitable and Accurate Classification of Tennis Skill Levels and Training Phases

Gyanna Gao[a], Hao-Yu Liao[b], Zhenhong Hu[c,d*]

[a] Eastside High School, Gainesville, FL.

[b] Department of Environmental Engineering Sciences, University of Florida, Gainesville, FL.

[c] University of Florida Intelligent Clinical Care Center, Gainesville, FL.

[d] Department of Medicine, Division of Nephrology, Hypertension, and Renal Transplantation, University of Florida, Gainesville, FL.

**\*Corresponding Author:** Zhenghong Hu, PhD, University of Florida Intelligent Clinical Care Center, Division of Nephrology, Hypertension, and Renal Transplantation, Department of Medicine, University of Florida, PO Box 100224, Gainesville, FL 32610-0224. Email: zhenhong.hu@medicine.ufl.edu

# Abstract

Numerous studies have demonstrated the manifold benefits of tennis, such as increasing overall physical and mental health. Unfortunately, many children and youth from low-income families are unable to engage in this sport mainly due to financial constraints such as private lesson expenses as well as logistical concerns to and back from such lessons and clinics. While several tennis self-training systems exist, they are often tailored for professionals and are prohibitively expensive. The present study aims to classify tennis players' skill levels and classify tennis strokes into phases characterized by motion attributes for a future development of an AI-based tennis self-training model for affordable and convenient applications running on devices used in daily life such as an iPhone or an Apple Watch™ for tennis skill improvement. We collected motion data, including Motion Yaw, Roll and Pitch from inertial measurement units (IMUs) worn by participating junior tennis players. For this pilot study, data from twelve participants were processed using Support Vector Machine (SVM) algorithms. The SVM models demonstrated an overall accuracy of 77% in classifying players as beginners or intermediates, with low rates of false positives and false negatives, effectively distinguishing skill levels. Additionally, the tennis swings were successfully classified into five phases based on the collected motion data. These findings indicate that SVM-based classification can be a reliable foundation for developing an equitable and accessible AI-driven tennis training system.

**Key Words:** Tennis training, AI, SVM, Classification, Tennis swing phase classification

# 1 Introduction

Research indicates that tennis, a lifelong activity, significantly benefits both teens and adults by reducing body fat percentage, lowering the risk of heart disease, and enhancing bone and overall physical health (Spring, Holmes, and Smith 2020; Pluim et al. 2007). Additionally, tennis improves psychological well-being, reduces depression and anxiety symptoms (Yazici et al. 2016), and offers cognitive and personality benefits among youth such as increased executive function (Xu et al. 2022) and more openness toward new experiences, creativity and flexibility (Demir et al. 2016). The lifelong effects of tennis, such as lowering obesity and heart disease levels (Spring, Holmes, and Smith 2020), underscore its importance. Unfortunately, many teens from low-income families cannot afford to participate in this sport due to the high costs of coaching, training, and transportation (Pandya 2021).

Tennis coaches typically emphasize correct body movement and stroke technique through demonstrations and drills. The forehand swing is the most fundamental and commonly used swing in recreational and competitive match play (Giménez-Egido et al. 2020; Myers et al. 2019). Correct execution of the forehand swing involves distinct phases that maximize performance and minimize injury risk (May B. M. 2020; Oshita et al. 2019). Intermediate-level players generally exhibit more consistent and standardized swing strokes, contributing to their skill level and reducing the risk of injury. Developing an affordable self-training system could function as a virtual coach, providing feedback to improve tennis skills and making the sport more accessible, especially for those from financially disadvantaged backgrounds.

In addition to the intuitive feedback from coaches for skill improvement, video recording and playback mechanisms have been used for performance analysis and evaluation for further improvement, especially for advancing tennis players. This process is more objective and nuanced than the coaches' intuitive feedback, but more time-consuming with the need for additional analysis of the video data. For instance, Li (2022) used computer, camera, and video analysis techniques to obtain data for tennis training and teaching of serving techniques among adult tennis professionals and found significant differences between the experimental group and the control group, with and without video analysis, respectively. Considering the additional cost and time in conducting video analysis, it is not a feasible option for affordable self-training for disadvantaged young tennis players. Oshita et al. (2019) developed a self-training system with a remote controller for a software connected to the optical motion capture via 12 cameras to allow visualization of real-time motion features of the beginner tennis learner compared to that of a skilled tennis player on a large screen for skill imitation and enhancement, an approach too costly and not feasible for disadvantaged tennis learners either.

To enhance the accessibility of tennis for young people from economically disadvantaged backgrounds, it is imperative to develop an application that operates on widely used and affordable devices such as smartphones and smartwatches. Such an application could empower underprivileged youth to independently cultivate and refine their tennis abilities without the financial burden of hiring personal coaches or attending regular clinics. However, this requires robust machine learning models to classify tennis skill levels and differentiate the tennis strokes into phases such as backswing, backloop, forward swing, follow-through, and recovery phases (May B. M. 2020).

Several commercial self-training systems, such as the ZEPP® tennis swing analyzer from Zepp US, Inc. (Giménez-Egido et al. 2020), Babolat Play, and the SMART TENNIS SENSOR from Sony (STSS) (Büthe et al. 2016), have been developed to analyze the performance of tennis players. However, these systems are expensive and often lack accuracy and precision in feedback. They also require specific hardware, limiting their accessibility. For example, while the ZEPP swing analyzer (Giménez-Egido et al. 2020), Babolat Play and STSS (Büthe et al. 2016) all need to install a sensor to the racket, only certain racket brands are compatible with STSS. Rigozzi et al. (2023) (Rigozzi, Vio, and Poronnik 2023) developed and tested the Tennis Racket Accelerometer MyoWare Wearable Device Version 2 (Tram-2), which has eight parts that are to be attached to the different spots of the racket handle as well as the wrist and elbow for data collection, which is not feasible and convenient. No studies have been done on the accuracy of these sensors in skill level classification and tennis swing phase classification. While ZEPP can be used for any racket, commercial review on it has been unfavorable and inconvenient. The only study with ZEPP (Giménez-Egido et al. 2020) assumed its accuracy, utility, reported number of ball strikes, ball impact and spin, racket head speed, match time, and calories burned descriptively without any information on the differences in skill levels and swing accuracies and with unknown accuracy of its sensor data. Myers et al. (2019) utilized STSS to measure the hitting volume accuracy of advanced high school tennis players across different tennis swings and only generated descriptive information on players of number of uses of forehand, backhand, and serves over a span of six weeks' time. This descriptive performance analysis does not allow for evaluative feedback for tennis players in skill improvement.

Existing self-training systems suffer from three primary issues: the complexity and high cost of data collection systems, the limited focus on professional players, and the lack of accessible solutions for beginners and youth players. Additionally, to the best of our knowledge, no study has yet utilized the built-in sensors of the Apple Watch™ to collect motion data for potential tennis self-training applications.

The first goal of this study was to validate the utility of the Apple Watch™ for motion data collection, informed by Espinosa et al. (2020), who confirmed the accuracy of its inertial and heart rate sensors compared to commercial-grade devices. The second goal was to develop machine learning models based on motion data from the Apple Watch™ to classify the skill levels of junior tennis players and classify tennis strokes into phases. The study aims to build an AI-based tennis self-training model for affordable and convenient use on everyday devices such as iPhone and Apple Watch™.

Introducing machine learning for motion analysis in tennis training is transformative. This AI application fills a crucial market gap by offering personalized insights through comprehensive data analysis, allowing users to enhance their skills with tailored assessments. The system facilitates real-time data collection using accessible and cost-effective devices, making advanced training capabilities available to a broader demographic. The machine learning-based motion analysis and training system, implemented using Support Vector Machine (SVM) models, represents a significant advancement in sports technology, particularly for tennis training.

# 2 Materials and Methods

## 2.1 Participants

Twelve minor participants were recruited with parental consent. Each participant completed a background survey detailing their demographic information and tennis skill levels. The gender distribution was 4 males and 8 females. The skill levels were evenly split, with 6 beginners and 6 intermediate players. The participants included 3 African Americans, 5 Asians, and 4 Whites. Two participants were left-handed. Ages ranged from 8 to 16 years.

## 2.2 Data Collection

The on-court motion data collection sessions for this research project were conducted at a local tennis center. Participants were instructed to stand at the intersection of the service lines and hit normal-paced balls aimed at the middle of the opposite court. They completed five sessions: the first and second sessions involved hitting 10 forehands each, the third session involved hitting 20 forehands, and the fourth and fifth sessions involved hitting 10 backhands each. Participants wore an Apple Watch™ on their dominant hand. Participants warmed up their forehand and backhand groundstrokes with hand-fed balls. The experiment commenced once they felt ready, with pre-counted balls and a designated area circled on the court to ensure consistent ball feeding height and location.

Participants wore the same Apple Watch™ on their wrist throughout the sessions. The Apple Watch™'s Inertial Measurement Units (IMUs), which included accelerometer and gyroscope sensors, measured attributes such as specific force, acceleration, change in velocity over time, and angular motion rate during tennis racket swings. A third-party application, SensorLog™, running on the Apple Watch™, captured the motion-related features, which was then transferred to a personal device for analysis. According to prior research (Susta and Connell 2009), the wrist angle is a crucial factor in a player's tennis performance. This study focused on three motion features: Motion Yaw, Roll, and Pitch, collectively known as the Euler angles, which measure the orientation of 3D motion—in this case, the tennis swing. Therefore, these three features were selected for the study and given the importance of forehand swings, the forehand swing data were used for analysis and model building.

## 2.3 Data Preprocessing

The collected data was preprocessed to remove noise and handle missing values. Additionally, data transformation steps, including standardization and normalization of the Motion Yaw, Roll, and Pitch features, were performed to ensure consistency in range. The ML-based tennis skill analysis system comprises two main modules: a self-evaluation module for assessing and classifying the skill levels of the tennis player, and a self-analysis module for automatically classifying larger movement units (e.g., the entire tennis stroke) into five distinct phases, as illustrated in Figure 1 below.

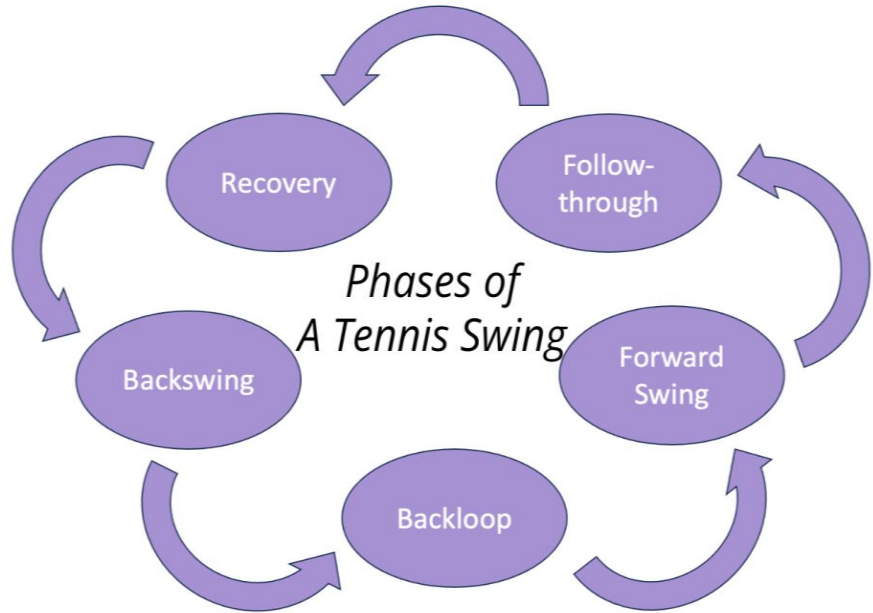

Figure 1. Phases of a tennis swing (adaption from May (2020))

The graph in Figure 2 below illustrates the swing trajectory of the first recorded tennis swing and hit, highlighting three key motion features—Motion Yaw, Roll, and Pitch—captured by the Apple Watch™.

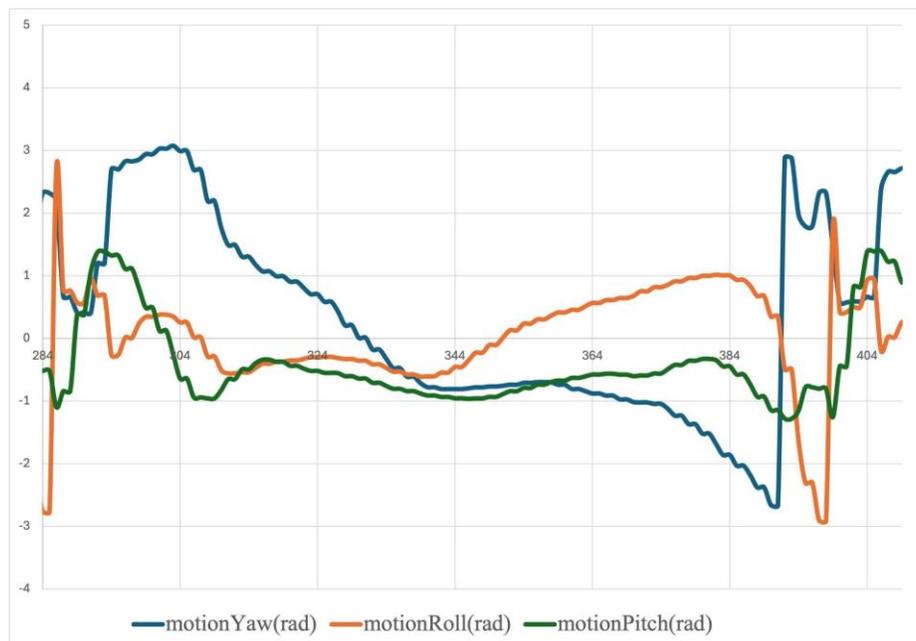

Figure 2. First tennis hit motion Yaw, Roll, Pitch

PCA was utilized to reduce the dimensionality of the data while retaining the most significant variance in the dataset. This step was crucial to simplify the subsequent analysis by focusing on the most informative aspects of the motion data. By transforming the original features into a set of uncorrelated principal components, PCA helps in mitigating noise and redundancy, thereby enhancing the efficiency and accuracy of the change point detection process conducted by the Ruptures library. Details of offline change point detection methods can be found in (Truong, Oudre, and Vayatis 2020). Figure 3 presents a close-up analysis of one swing stroke segmented into five phases, with the numbered ticks on the horizontal axis representing the number of frames.

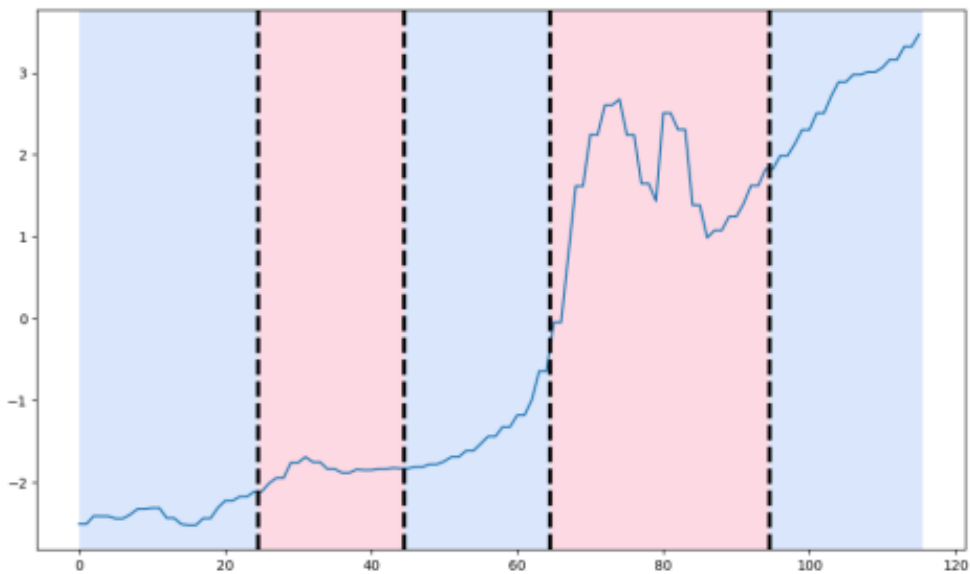

Figure 3. A close-up shot analysis of one swing stroke segmented into five phases with the numbered ticks on the horizontal line representing the number of frames.

In the initial phase, the player executes a backswing by turning and lifting their racket to the side, signifying the start of the swing motion in anticipation of the ball. The next phase involves dropping the racket head downward with the help of gravity, contributing to the topspin generated upon contact with the ball. The player then transitions from the backswing to the forward swing, moving the racket upward to contact the ball. The rapid motion of the racket making contact with the ball occurs between the forward swing and follow-through phases, so it is not labeled as a separate phase.

In the follow-through phase, the player continues the swing after making contact with the ball to generate spin and power, directing the ball to the intended placement. Finally, the player moves the racket behind the shoulder of their non-dominant hand, allowing for smooth repositioning of the racket to prepare for the next shot.

## 2.4 SVM Classification

In this study, we employed Support Vector Machine (SVM) binary classification to automatically classify participants' tennis performance levels into either beginner or intermediate categories. The SVM algorithm was selected for its robustness in handling high-dimensional data and its effectiveness in finding the optimal hyperplane that separates the two classes with maximum margin (Ritu Sharma, Kavya Sharma, and Apurva Khanna 2020). SVM is known for its ability to generalize well to unseen data, making it less prone to overfitting compared to some other methods. Additionally, SVM models are resilient to outliers in the training data, making them suitable for small-sample classification (Myers et al. 2019). To automatically classify tennis swing phases, we developed a multi-class classification SVM model.

Figure 4 provides a flowchart of the experimental design, offering an overview of the study. The dataset was divided into 80% for training and validation and 20% for testing. To address the challenge of limited training data, we adopted 5-fold cross-validation to ensure sufficient samples for both training and validation (Figure 5). In our study, we used SVM models with three different kernels to evaluate their performance in classifying tennis skill levels: Radial Basis Function (RBF), polynomial (Poly), and sigmoid. Each kernel's performance was assessed based on the averaged 5-fold cross-validation metrics, including accuracy, precision, recall, and F1 score, to ensure their generalizability and accuracy in classifying new, unseen data. This approach provided a comprehensive evaluation of each kernel's effectiveness in the context of our research.

For phase segmentation, we developed a self-analysis module that leveraged the Rupture library, a robust Python tool specifically designed for detecting changepoints (signal shifts) in time series data (Truong, Oudre, and Vayatis 2020). The Rupture library facilitated the identification of significant transitions within the motion data, allowing for the precise segmentation of larger movement units into distinct phases. Following the changepoint detection, we implemented a multi-class SVM model to automatically classify the segmented movement units into five predefined phases. These phases represent the various stages of a tennis swing, including backswing, backloop, forward swing, follow-through, and recovery.

To assess the performance of the multi-class SVM model, we employed the Receiver Operating Characteristic Area Under the Curve (ROC-AUC) metric. ROC-AUC is a comprehensive performance measure that evaluates the model's ability to distinguish between different classes. By analyzing the ROC-AUC, we ensured that the multi-class SVM model not only accurately classified the phases but also maintained a high level of discrimination between them. This rigorous evaluation process provided confidence in the model's capability to reliably segment and classify tennis swing phases, contributing to the development of an effective AI-driven tennis training system.

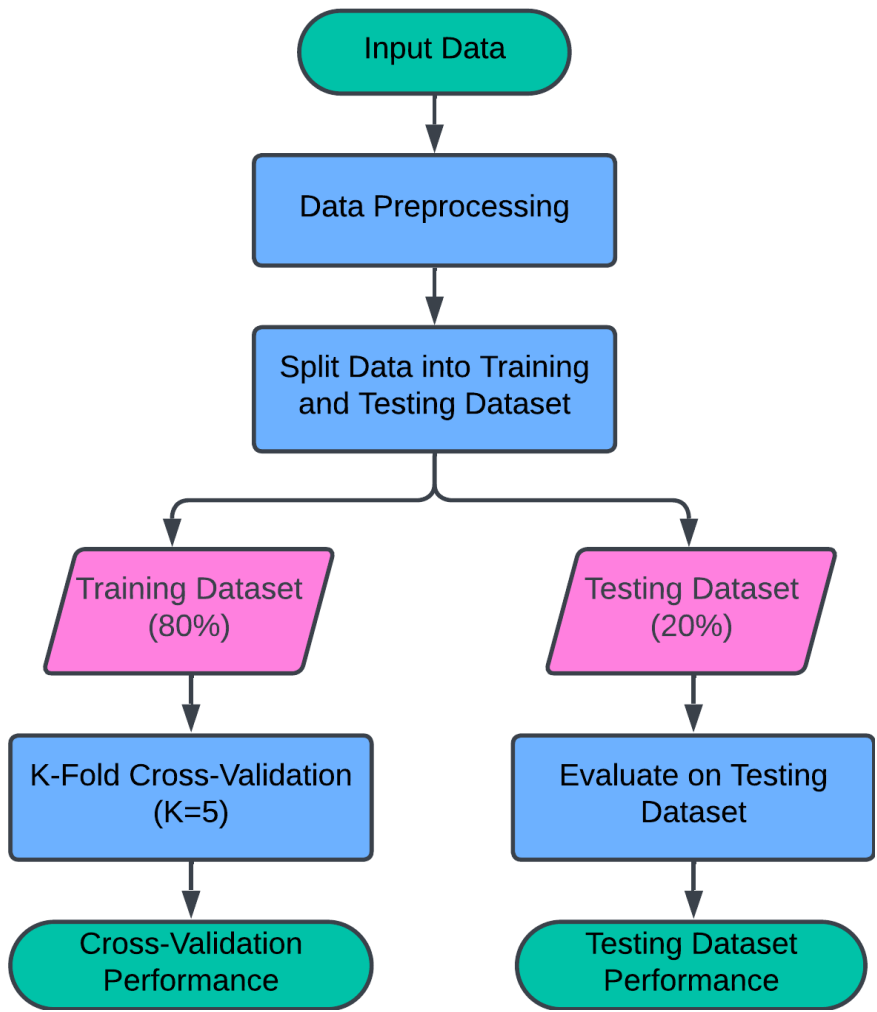

Figure 4. The flowchart of the experimental design

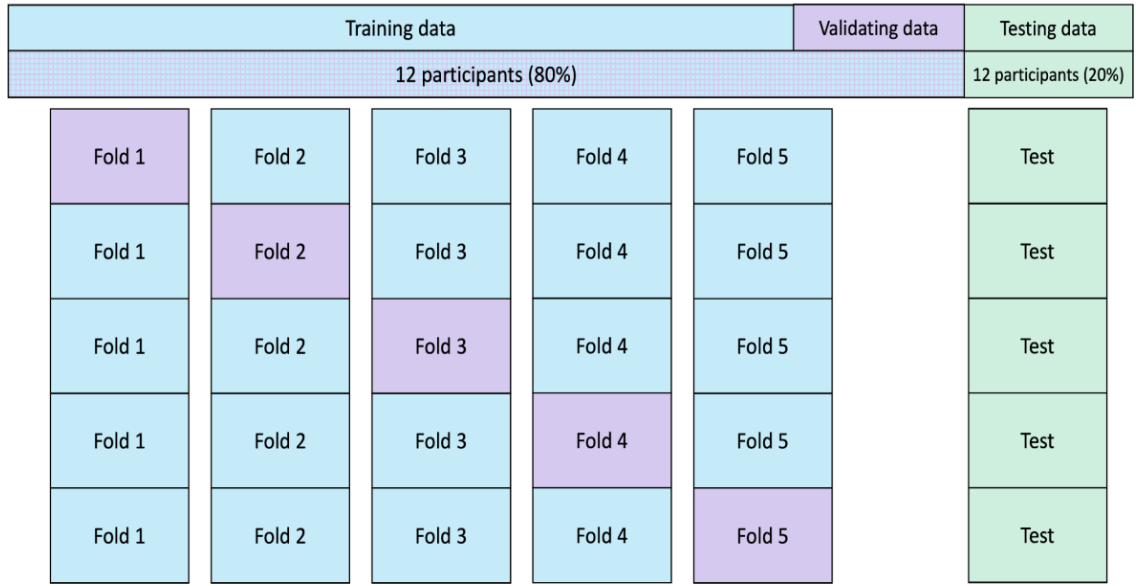

Figure 5. The 5-fold cross-validation on 12 participants

## 2.5 Ethics Statement

Institutional Review Board (IRB) approval was obtained from the Alachua County School Board, and permission was granted for the use of the tennis courts for data collection. Informed consent was secured from all minor participants and their parents. To ensure confidentiality, participants' names were collected but were de-identified using numerical IDs for analysis purposes. The data was securely stored on a password-protected personal computer for both collection and analysis.

# 3 Results

## 3.1 Descriptive Analysis

The analysis revealed noticeable visual differences in the tennis swings of participants at beginner and intermediate levels. These differences are illustrated in Figures 6. Figure 6A depicts the swing trajectory of Participant 1011, a beginner-level player. The three consecutive peaks observed in this figure show varied trajectories, appearing choppy and lacking fluidity. This irregular pattern is typical of beginner swings, where the lack of consistent practice results in uneven and less controlled movements. The variability in these peaks indicates that the player is still developing muscle memory and the coordination required for a smooth, consistent swing. Figure 6B, on the other hand, illustrates the swing strokes of Participant 1012, an intermediate-level player. In this figure, the peaks are more uniform and consistent, demonstrating a smoother and more controlled swing. This consistency is a hallmark of intermediate-level players, who typically have more experience and practice, leading to a more refined technique. The similarity in the trajectories of these peaks suggests that the player has developed better coordination and muscle memory, allowing for a more reliable and repeatable swing motion.

These visual differences highlight the distinct characteristics of tennis swings at different skill levels. Beginner players tend to have more erratic and less polished swings, while intermediate players exhibit more consistency and control in their movements. This distinction underscores the potential of using motion data analysis to accurately classify and assess the skill levels of tennis players, providing valuable insights for developing personalized training programs.

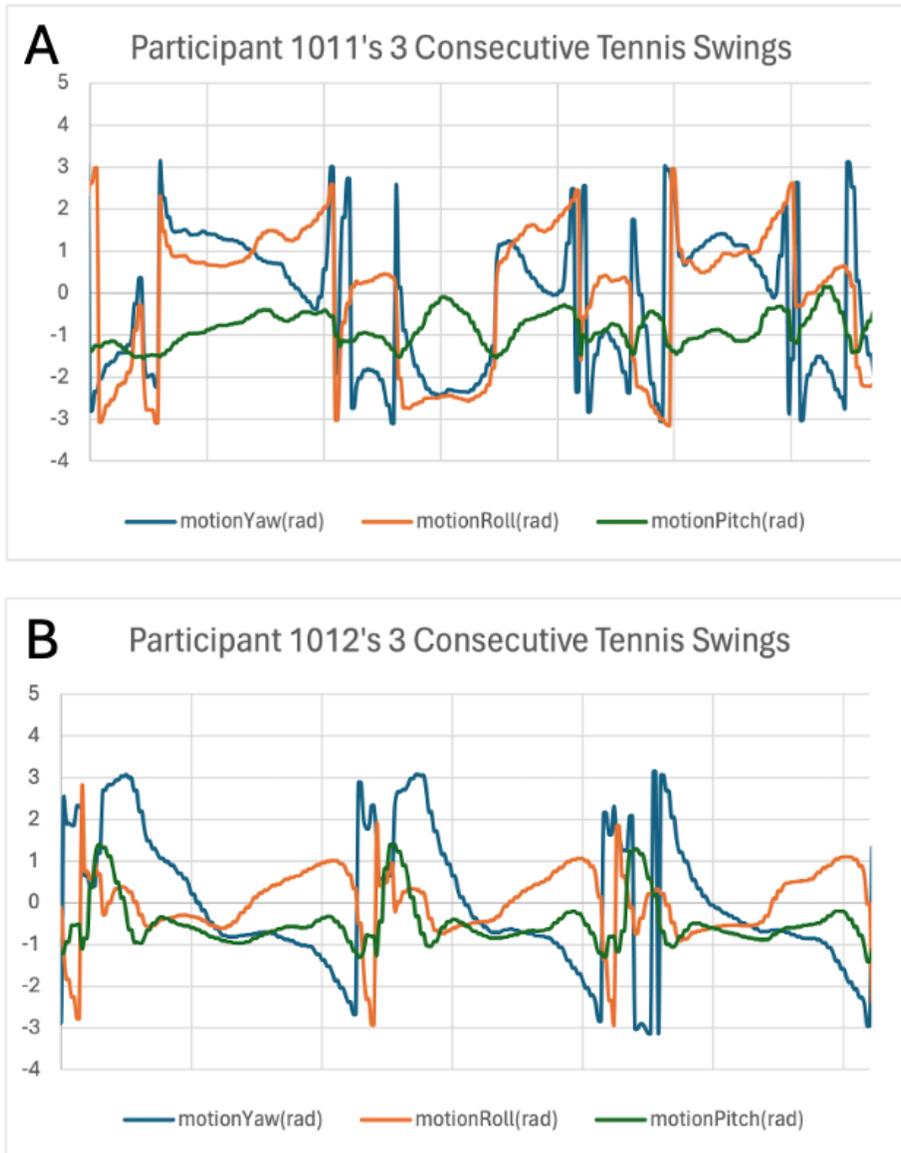

Figure 6. Three different but consecutive peaks in one beginner tennis (A) and intermediate-Level player's strokes

## 3.2 Participant Performance Level: Binary Classification

The performance of the SVM binary classification model in distinguishing between beginner and intermediate tennis players was assessed using three different kernels: Radial Basis Function (RBF), polynomial (Poly), and sigmoid. The evaluation process involved both cross-validation and testing phases to ensure the robustness and generalizability of the models (Table 1).

During the cross-validation phase, the RBF kernel emerged as the most effective, demonstrating superior performance across several metrics. It achieved the highest accuracy, indicating its ability to correctly classify the majority of participants. Additionally, the RBF kernel showed balanced precision and recall scores, suggesting that it was effective at identifying both true positives and true negatives while minimizing false classifications. This balance is crucial for real-world applications, where both types of errors can have significant implications.

In contrast, the polynomial (Poly) kernel, despite its high precision, exhibited a significantly lower recall. This imbalance indicates that while the Poly kernel was highly accurate in its positive classifications, it failed to identify a substantial number of true positive cases, leading to a lower overall accuracy. The sigmoid kernel showed performance metrics similar to the Poly kernel, further highlighting its inadequacy for this classification task. The lower precision and recall scores of the sigmoid kernel suggest that it struggled with both identifying true positives and avoiding false positives.

During the testing phase, the RBF kernel maintained its superior performance, confirming its reliability in real-world scenarios. The consistency of the RBF kernel's metrics between cross-validation and testing phases underscores its robustness and ability to generalize well to new, unseen data. This makes the RBF kernel a highly reliable choice for developing an AI-driven tennis training system aimed at accurately assessing and improving players' skills.

Table 1. Model performance evaluation using 3 different kernels: Radial Basis Function (RBF), polynomial (Poly), and sigmoid kernels

| Evaluation | Kernel | Accuracy | Precision | Recall | F1 Score |
|---|---|---|---|---|---|
| **Cross-Validation** | **RBF** | **0.772** | **0.742** | **0.750** | **0.746** |
| | Poly | 0.509 | 0.926 | 0.477 | 0.630 |
| | Sigmoid | 0.511 | 0.458 | 0.458 | 0.458 |
| **Testing** | **RBF** | **0.771** | **0.743** | **0.748** | **0.745** |
| | Poly | 0.508 | 0.927 | 0.477 | 0.630 |
| | Sigmoid | 0.509 | 0.459 | 0.456 | 0.457 |

The detailed evaluation provided by the confusion matrix (Figure 7) further illustrates the effectiveness of the RBF kernel. The matrix reveals that the model accurately classified a significant number of beginners and intermediate players, with relatively low rates of false positives and false negatives. This high level of accuracy and balanced performance demonstrates the RBF kernel's capability to distinguish between different skill levels effectively.

Given the superior performance of the RBF kernel, it will be employed for subsequent multi-class SVM phase classification of tennis strokes. Its robustness in capturing intricate patterns and handling non-linear separations in data will enhance the accuracy and reliability of phase segmentation. This will facilitate a more detailed and nuanced analysis of tennis performance across different skill levels, providing valuable insights for personalized training programs. The use of the RBF kernel in multi-class classification underscores its pivotal role in advancing AI-driven methodologies in sports training and performance analysis.

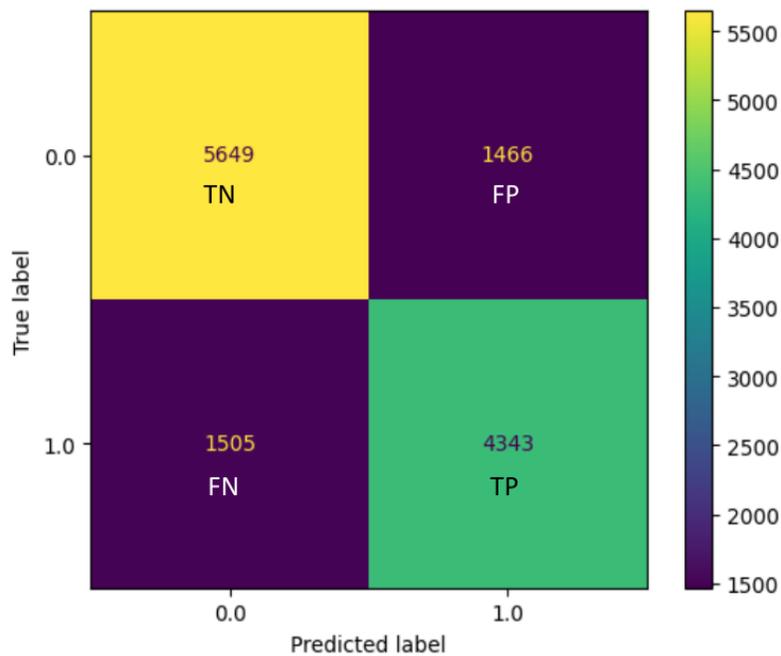

Figure 7. Confusion matrix of testing results using the RBF kernel for skill level classification (0: Beginner level; 1: Intermediate level)

# 3.3 Tennis Strokes Phase Classification: Multi-Class Classification

The multi-class classification of tennis stroke phases using a multi-class SVM model was evaluated, and the results are summarized in the average confusion matrix in Figure 8. This matrix reveals that the model's performance varied across the five phases. For Phase 1, while the model correctly identified a significant number of instances, it also showed considerable misclassifications into Phase 2, suggesting a need for further differentiation between these phases. Phase 2 displayed similar challenges, with a substantial overlap with Phase 1, indicating that their distinguishing features are not sufficiently distinct. Phase 3 had notable misclassifications spread across multiple phases, highlighting less clearly defined characteristics for this phase. Phase 4, despite a fair number of correct classifications, had high misclassifications into Phase 5, suggesting shared common features that need more precise definition. Phase 5, although performing relatively well, also showed significant overlap with Phase 4, underscoring the necessity for further refinement. These results indicate that while Phases 1 and 5 had the highest correct classifications, significant misclassifications across all phases point to the need for further optimization to enhance the model's discriminative power.

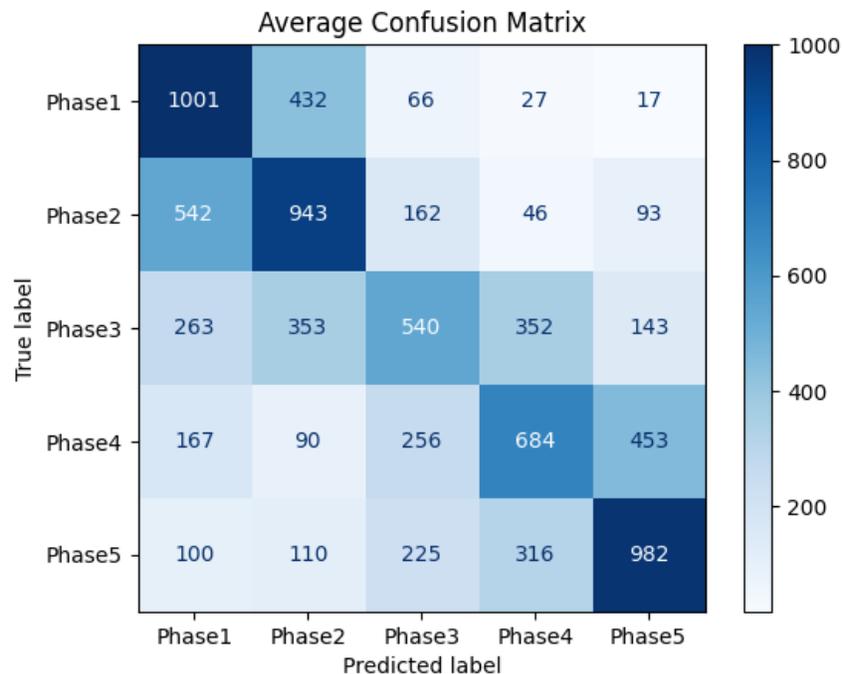

Figure 8. Average confusion matrix for tennis stroke phase classification using SVM multi-class classification

The Receiver Operating Characteristic (ROC) curves in Figure 9 further evaluated the model's performance, with Class 0 achieving the highest AUC of 0.87, demonstrating excellent discrimination capability. Classes 4 and 1 followed with strong performance, but Classes 3 and 2 had lower AUC values, indicating difficulties in accurately distinguishing these phases. Overall, while the SVM model shows promise, particularly for certain phases, additional refinement in feature extraction and model complexity is required to improve the classification performance across all phases.

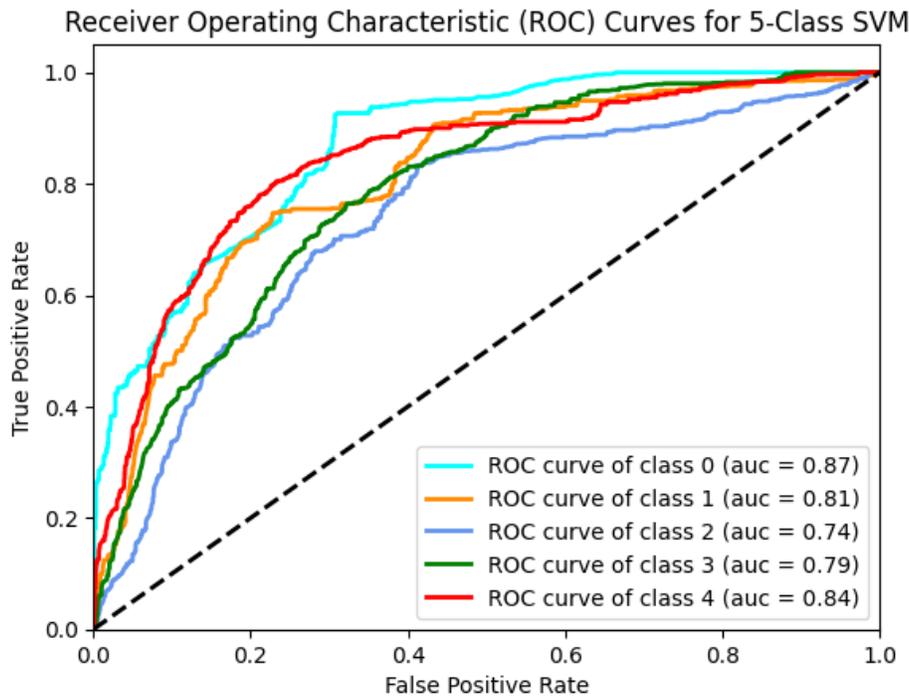

Figure 9: Receiver Operating Characteristic (ROC) curves for the 5-class SVM multi-class classification of tennis stroke phases, with Area Under the Curve (AUC) values for each class

# 4 Discussion

This study demonstrated that our binary classification algorithm successfully differentiated between beginner and more advanced tennis players with an accuracy of 77%. Additionally, our multi-class classification algorithm effectively distinguished the five major phases of tennis swings, achieving ROC-AUC values ranging from 0.74 to 0.87. The accuracy rates of these models provide promising evidence for their future application in developing affordable and accessible mobile applications, which could help make tennis a more equitable sport, particularly for individuals from economically disadvantaged backgrounds.

One of the unique aspects of this study is its reliance on real-world data collected from a diverse group of children and youth, encompassing various ethnic and gender backgrounds as well as different levels of tennis skill. Unlike previous studies that used existing datasets unrelated to tennis, such as Song (2024), or those that focused primarily on competitive players Zhang (2024) and adults Li (2022), our research specifically targets a more inclusive demographic.

Furthermore, our results highlight the feasibility of using Apple Watch™ sensors combined with the SensorLog app to collect motion data. This method proved to be more convenient than traditional systems that rely on cameras and video analysis techniques, such as those used by Li (2022). By leveraging the widespread accessibility of devices like the Apple Watch™, our approach can significantly lower the barriers to entry for tennis training.

The potential impact of this technology extends beyond competitive play to recreational use, particularly for children and youth from financially disadvantaged backgrounds. Tennis offers numerous health benefits, including reduced body fat percentage, improved bone health, and enhanced overall physical and mental health (Pluim et al. 2007). However, the high costs associated with private coaching and group clinics often make it inaccessible to low-income families (Pandya 2021). The robust algorithms we developed, combined with the convenience of using an Apple Watch™, pave the way for an AI-facilitated tennis training app that can provide affordable and accessible training. Such an app would support skill development and enjoyment of tennis, contributing to greater equity in the sport and promoting both physical and mental health among disadvantaged youth.

Despite these promising results, our study has several limitations. The participant sample size was small, although the sensors collected a vast amount of data frames. Increasing the number of participants would enhance the training and testing of our algorithms, potentially improving accuracy. Additionally, we selected only three motion data features—motionYaw, motionRaw, and motionPitch—out of the 51 variables captured by the Apple Watch™. While these Euler angles provide a comprehensive evaluation of tennis swing orientation and direction, incorporating other features like location and accelerometer data could further refine the algorithms.

Moreover, our models used fixed C and Gamma parameters for the different kernels, which may limit their performance. Fine-tuning these parameters could enhance the model's accuracy. Although the SVM model proved effective, it is inherently a binary classification tool. More advanced AI models, such as Long Short-Term Memory (LSTM) networks and Temporal Convolutional Network (TCN), could offer improved multi-classification motion data analysis by handling sequential data and capturing long-term dependencies. Comparing the effectiveness of LSTM, TCN and SVM models could provide insights into which model offers superior predictability and resilience against overfitting. Future research should explore these advanced models to further enhance the robustness and accuracy of the algorithms.

# 5 Conclusion

The SVM models developed in this study achieved an overall accuracy of 77% in classifying tennis players as beginners or intermediates, demonstrating their effectiveness in distinguishing between different skill levels with low rates of false positives and false negatives. Furthermore, the classification of tennis swings into five distinct phases based on the collected motion data underscores the potential of SVM-based classification as a reliable foundation for an AI-driven tennis training system. These findings support the future development of an AI-based tennis self-training model that is both affordable and convenient, leveraging everyday devices such as the iPhone and Apple Watch™ to enhance tennis skills.

Future research can extend this study in several ways. Firstly, while our current analysis focuses on beginner and intermediate levels, including professional players could further enhance the training goals. Secondly, exploring the integration of imaging technology with the Apple Watch™ application could provide a more comprehensive and cost-effective training solution by combining smartphone video recording with motion data. Lastly, applying ensemble learning techniques could help reduce model variance and improve accuracy, thereby enhancing the robustness of future studies.

**Funding Information**

This research received no external funding.

**Data Availability**

The data supporting the findings of this study is available upon reasonable request. Researchers interested in accessing the data are encouraged to contact the corresponding author, including a brief description of the research purpose and intended use of the data. The data will be provided in a format consistent with ethical guidelines and privacy policies governing the study. Additionally, supplementary information that may assist in understanding and analyzing the data can also be made available upon request. For transparency and reproducibility, we aim to facilitate data sharing while ensuring compliance with relevant data protection regulations.

## Declarations

**Conflict of Interest**

No potential conflict of interest was reported by the authors.